# A programmable clock generator for automatic Quality Assurance of LOCx2

Zhi-yue Wang, Tian-kuan Liu, Qi-jie Tang, Yi Feng, Jian Wang, *Senior Member, IEEE*

*Abstract–.* **The upgrade of ATLAS Liquid Argon Calorimeter (LAr) Phase-1 trigger requires high-speed, low-latency data transmission to read out the Lar Trigger Digitizer Board (LTDB). A dual-channel transmitter ASIC LOCx2 have been designed and produced. In order to ensure all the LOCx2 chips behave properly, a Quality Assurance needs to be conducted before assembly. The problem I was trying to solve in this project is to yield a clock signal with continuously adjustable frequency and phase offset to generate and control an eye diagram for the QA. By configuring the registers of an any-frequency generator IC, Si5338, the clock signal whose frequency range from 5MHz to 200 MHz have been properly produced. For the purpose of further development, a C-language based DLL which packs up the function of adjusting frequency and setting phase offset was designed and built, and several evaluation was performed to ensure the robustness of DLL.**

## I. INTRODUCTION

The upgrade of ATLAS Liquid Argon Calorimeter (LAr) made a demand of a high-speed, low-latency data transmission to read out from the LAr Trigger Digtizer Board(LTDB). In the LAr system, the detector was connected to the front-End board, and a copy of raw wave-form collect from detector pass through a linear mixer and a Layer Sum Board (LSB). The output signal goes into the LTDB which split the signal into four ways of ADCs and the data yield by ADCs was caught by LOCx2, which was an ASIC designed to meet the requirement of high-speed and low-latency data transmission. This part of system was shown in the Fig.1.

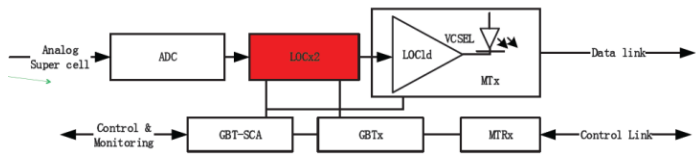

Fig. 1. Link system block diagram

About 7000 LOCx2 were fabricated, therefore a Quality Assurance need to be done in order to pick out those dies with better performance.

The amount of dies was too huge to perform a pure artificial test one by one, thus an automatic test system was designed, which perform tests include eye diagram, Phase-Lock loop ,Bit Error Rate(BER), I2C and input skew. Fig.2. shows the Block diagram of LOCx2 eye diagram and PLL test setup. All 7000 LOCx2 chips will test the serial output eye diagram. About 1% of the chips that pass the eye diagram tests and BER tests in each wafer will be sampled to measure the Phase-Locked-loop (PLL) tuning range, the jitter of the serial output.

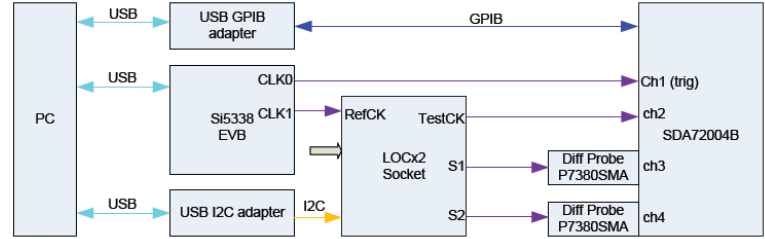

Fig.2 Block diagram of LOCx2 eye diagram and PLL test setup

Fig.3 shows the Block diagram of LOCx2 BER, I2C, and input skew test setup. All 7000 LOCx2 chips will get through the test of BER and I2C. About 1% of the chips that pass the eye diagram tests and BER tests in each wafer will be sampled to measure the 3.125-ns input skew tolerance in the ASIC ADC mode.

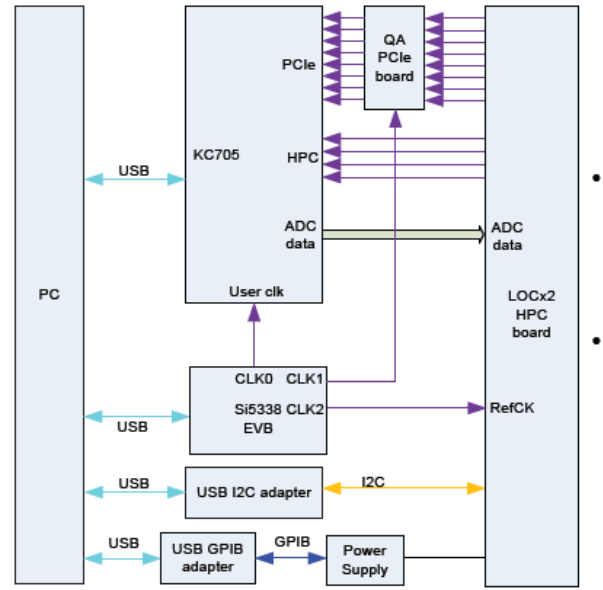

Fig. 3. Block diagram of LOCx2 BER, I2C, and input skew test setup

This work was supported by the National Natural Science Funds of China under Grant No: 11603023, 11773026, 11728509, the Fundamental Research Funds for the Central Universities (WK2360000003, WK2030040064), the Natural Science Funds of Anhui Province under Grant No: 1508085MA07, the Research Funds of the State Key Laboratory of Particle Detection and Electronics, the CAS Center for Excellence in Particle Physics, the Research Funds of Key Laboratory of Astronomical Optics & Technology, CAS.

The Authors Zhi-yue Wang, Qi-jie Tang, Yi Feng, Jian Wang are with the State Key Laboratory of Technologies of Particle Detection and Electronics, University of Science and Technology of China, Hefei, Anhui 230026, China (e-mail: Jian Wang, wangjian@ustc.edu.cn).

Tian-Kuan Liu is with the Department of Physics, Southern Methodist University, Dallas, TX 75275, USA, tliu@mail.smu.edu

As was shown in these two diagrams, Si5338 play the role of generating an any-frequency clock signal for test. A host PC will control and supervise the whole test system automatically.

The signal yielded by Si5338 was feed to the LOCx2 and a SDA72004B(an oscilloscope) simultaneously, and the output signal from LOCx2 together with the signal directly transfer into SDA72004B yield an eye diagram for test. And in the BER test in order to shorten the time, a test system which is able to test 6 dies simultaneously was produced and in this system, and the test signal was generated by KC705 which was also feed by Si5338, therefore the Si5338 need to have the ability of adjusting the frequency of clock signal continuously with programming interface. A further test of BER request an input signal with an alterable phase. All these function need to be provided by Si5338 EVB which does not provide a programming interface. So the main purpose of this work is to develop a set of Application Programming Interface (API) in the PC side for further development and a MCU firmware to control the Si5338 to generate up to four channels of any-frequency clock with controllable phase, which make it possible to control the entire system with programming automatically.

II. STRUCTURE OF SIGNAL GENERATION SYSTEM

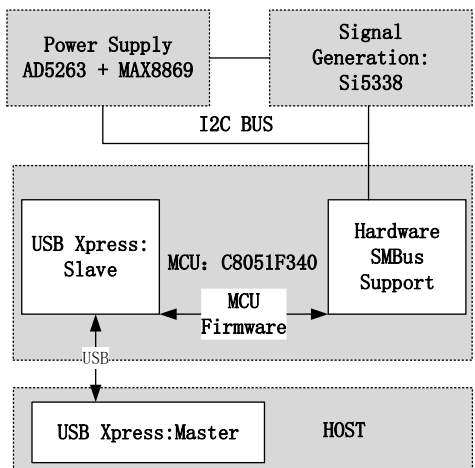

Fig. 4.Block Diagram of Si5338 EVB with host computer

As was shown in Fig.4, Si5338 is a programmable, any-frequency, four channels clock signal generator, whose output signals can be configured by its registers, and this is the kernel for generating the clock signals. C8051F340 is a MCU based on CIP-51 core which maintains the function of the hardware SMBus I/O, and an USB-controller, which is used to communicate with SI5338 through I2C bus to configure the output of clock signal and the voltage of power supply modules for Si5338. AD5263 is a programmable variable resistor and MAX8869 is a voltage regulator. In order to meet the power supply for five different VDD pins on the Si5338, five channels of power supply module which consist of an AD5263 and a MAX8869 were needed. AD5263 is controlled by MCU through I2C bus as well. All these components were on the Si5338 EVB board and make up the programmable clock generator.

Concerning the Si5338 EVB has a USB interface, and C8051F340 MCU has a USB controller built-in, The USB interface was choosen to send to and receive commands from MCU. A set of solution called USBXpress which could drive the USB interface in both host and device was adopted. Base on this software, we developed the host program as well as device firmware which communicate with each other through USB interface.

III. THE DESIGN OF DEVICE FIRMWARE

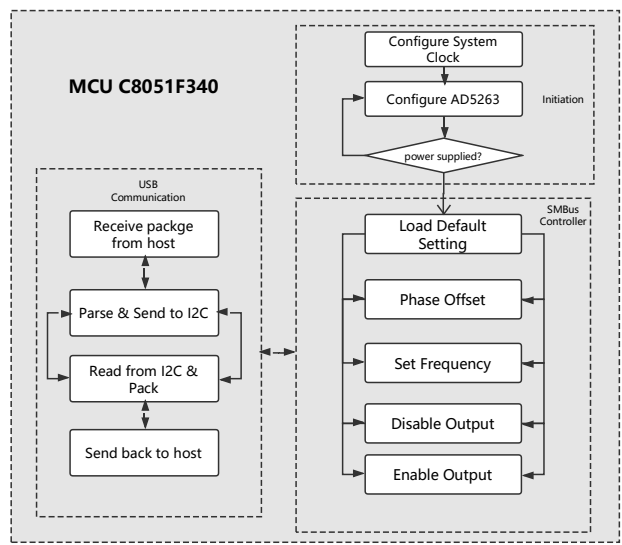

Fig. 5.Block Diagram of MCU Firmware

The C8051F340 has a firmware no larger than 4352Bytes, which make it difficult to integrate a complex set of decision program into MCU firmware after implemented the USB communication. Therefore the firmware is designed to provide a communication bridge from I2C to USB, as well as configure all five channels of the power supply module. Therefore, the application on the host can just call a set of routines which will send data about the device I2C address, register address and the value of register to device, and the value will be written to each register through I2C bus.

Fig.5 shows the structure of firmware, at the beginning of procedure, MCU would set up the system clock, register the interrupt vectors, configure USB and SMB controller and disable the hardware watch dog. Finished this basic startup routine, MCU would configure AD5263 and MAX8869 through I2C bus to set up the power supply for clock generator. After these procedure, MCU will enter a circulation in which MCU would judge if a set of symbol variable were set, if so MCU will perform the procedure indicated by symbol variable. These symbol variable was set in the interrupt routine of USB.

In the interrupt routine of USB which was trigged by receiving command from USB, device will firstly determine what kind of action was indicated by this command, read or write, by checking the first byte of command, as 0xFF represent write while 0x00 represent read.

If a 'write' action was to be performed, device will receive four bytes including the bytes of read or write. The second byte is the hardware I2C address of target device, the third byte is the address of register in target device, and the last byte is the value to write. In the USB interrupt routine, these three bytes will be record into three different variables, and FLAG_WRITE, the symbol variable would be set to 0xFF and the interrupt routine would finish. Back to main routine, program will catch this variable and write the address and value to the register SMB0DAT, followed by latching the bit5 of SMB0CN. MCU will enter the SMB interrupt in no longer than five times clock period after this procedure and perform a I2C writing. After returning to main process, the symbol variable will be cleared to 0x00. The whole process was shown in Fig.6

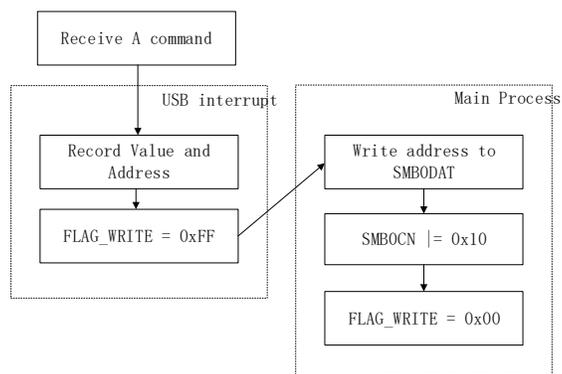

Fig.6. A Write Proccess using I2C interface

Similarly, if a 'read' action was to be performed, device will record the address of register to read, set the correlated symbol variable, and perform I2C reading. The different is that after all the process have been done, MCU will write the value of register back to host through USB. The read process was shown in Fig.7.

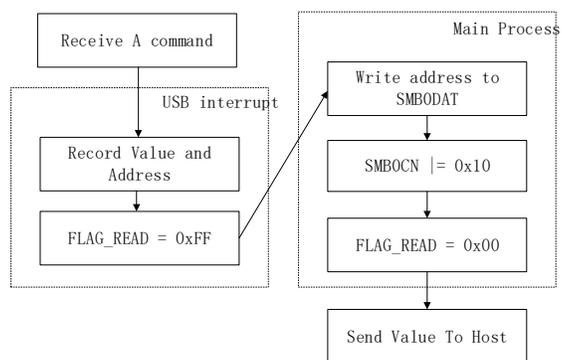

Fig.7.An Read Process using I2C interface

Since all the operation to Si5338 can be done by configuring the register through I2C bus, the function of firmware can be regard as simple as a USB to I2C adapter.

## IV. THE DESIGN OF HOST SOFTWARE

We try to decouple the host software into 4 layer, as shown in Fig.8, the fundamental layer drives the USB interface, and provide API for upper layer to read and write through USB. The second layer will pack up the protocol of command, which make it less complicated to instruct MCU. The third layer is the most important one, as all the configure processes were done in this layer, it calls the API provided by fundamental layer, and provide several API, including modifying the frequency of output clock signal, changing the phase and latching each channel of output. The third layer, application layer, is the user program which calls the API of second layer.

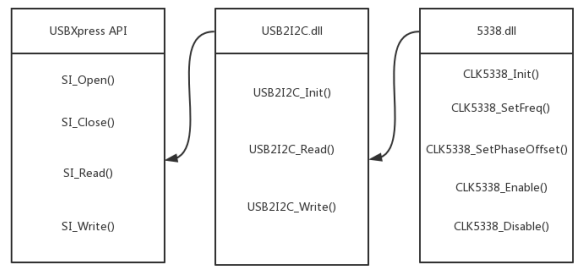

Fig.8. Layers of host software

The first layer mostly base on USBXpress, thus we can handle USB interface by calling a set of high-level API without dealing with USB protocol or developing system driver for USB by ourselves. The USBXpress provides a set of APIs, including SI_Open for establishing the USB communication, SI_Read for reading data from USB interface, SI_Write for writing data to USB interface, and USB_Close for terminating the USB communication.

USB2I2C.dll packs up these function into three APIs, each handle the read, write and initiation. When any of these APIs was called, USB2I2C.dll would call the interface of USBXpress and write command following the protocol mentioned in the previous section, and indicate the MCU to perform I2C read and write. It provides three interface, each handle the function of reading a value from a certain register, writing a value to a certain register, and initiation. This layer plays the role of middleware.

LIB5338.dll calls the API provided by USB2I2C.dll to configure the registers of Si5338, so as to change the frequency of output signals, adjust the phase offset, enable or disable a certain channel's output, and pack these function up to several APIs.

## V. CONCLUSION

We achieved the goal of generating an any-frequency signal and control the its phase offset, and we develop this set of software for both host PC and MCU, and the entire system works properly. Now the device have played its role in the whole testing system.

## VI. REFERENCES

[1] SILICON, L. "USB Xpress Programmer's Guide." (2007).

[2] Xiao, L., et al. "LOCx2, a low-latency, low-overhead, $2\times$ 5.12-Gbps transmitter ASIC for the ATLAS Liquid Argon Calorimeter trigger upgrade." Journal of Instrumentation 11.02 (2016): C02013.